\documentclass{article}
\usepackage{graphicx,ae}
\usepackage{epsfig}
\usepackage[T1]{fontenc}
\usepackage[latin1]{inputenc}
\usepackage{url}
\usepackage[]{hyperref}
\title{A SIMPLE MODEL FOR THE EVOLUTION OF EVOLUTION}
\author{\\Siegfried Fussy, Gerhard Gr\"ossing
and Herbert Schwabl\\
[.5cm]
{\small\em Austrian Institute for Nonlinear Studies,}\\
{\small\em Parkgasse 9, A-1030 Vienna, Austria}\\
{\small \url{http://web.telekabel.at/ains}}}%\\} %[1.5cm]}

\newcommand{\de}{\delta}
\newcommand{\ep}{\epsilon}

\newcommand{\bea}{\begin{eqnarray}}
\newcommand{\eea}{\end{eqnarray}}
\newcommand{\beq}{\begin{equation}}
\newcommand{\eeq}{\end{equation}}
\newcommand{\ba}{\begin{array}}
\newcommand{\ea}{\end{array}}
\newcommand{\bd}{\begin{displaymath}} % Glg. ohne Numerierung!!
\newcommand{\ed}{\end{displaymath}}
\begin{document}
\date{}
\maketitle
\vspace{.5cm}
\begin{abstract}
A simple model of macroevolution is proposed
exhibiting both the property of 
punctuated equilibrium and the dynamics of potentialities for different species
to evolve towards increasingly 
higher complexity. It is based on the phenomenon of 
fractal evolution which has been shown to constitute a fundamental property
of nonlinear discretized systems with one memory- or random-based 
feedback loop. The latter involves a basic ``cognitive" function of each
species given by the power of distinction of states within some predefined
resolution. The introduction of a realistic background noise
limiting the range of the feedback operation yields a pattern
signature in %one-dimensional 
fitness space with a distribution of temporal
boost/mutation distances
according to a randomized devil's staircase function.

Introducing a further level in the hierarchy of the system's rules, the
possibility of an adaptive evolutionary change of the resolution
itself is implemented, thereby 
providing a time-dependent measure of the species'
cognitive abilities: an additional feedback loop makes use of the inevitable
intrinsic fluctuations within the fitness landscape to direct the temporal
change of the resolution. Feeding back the small adaptive changes of resolution
into the essentially directionless variations of the patterns' lifetimes
in fitness space effectively leads to a clear tendency towards increasing
evolution potentials for each species ("hierarchically emergent fractal
evolution"). \\

\end{abstract}
%********************************************************************1
\section{Introduction: Punctuated equilibrium and self\--or\-gan\-i\-zed 
crit\-icality}

Punctuated equilibrium is a well-documented extension of Darwinian evolutionary
theory \cite{1goueld,2goueld}. 
On a population level, it has recently been observed in
the laboratory: during its evolution, the average cell size of
{\em Escherichia coli} bacteria showed
long periods of stasis with rare beneficial mutations rapidly sweeping through
the whole population \cite{1ele,lens}. Presently, however, it is not yet clear,
whether punctuations in the domain of populations and species, respectively,
are unrelated, or consequences of the same evolutionary mechanisms at work.
On the species level, some of the qualitative features of punctuated 
equilibrium have been successfully simulated by Bak and Sneppen 
\cite{1baks2} in a toy model on a lattice of sites, with each site
representing a species, thereby drawing on earlier work on self-organized
criticality (SOC) \cite{1baks1}. Recently, exact results have been obtained for
spatiotemporal evolutions in an extended SOC model of punctuated equilibrium
with many internal degrees of freedom per site \cite{1baks3},
for a model with neural-like rules \cite{sole1}, and for generating
phylogenetic-like trees \cite{vand}. However, the
successful implementation of SOC models notwithstanding, there exists one
important question which is hardly touched upon by these models, namely how
evolution produces entities of increased complexity. In particular, it has
been proposed by Kauffman and coworkers that the Darwinian process of mutation
and environmental selection would have to be complemented by self-organizing
processes in order to eventually decouple the strict %causal
link between
mutation and selection \cite{1kauff}.

In this paper we present a simple toy-model of evolution which is general
enough to be compatible with SOC models of punctuated equilibrium, but in
essential ways goes beyond them to encompass evolution towards higher
complexity. For, if one considers the Bak-Sneppen model and its more recent
versions, one is led to conclude that the SOC state is always a ``quasi - steady
state", i.e., ``the evolution process always self-organizes into the same
critical steady state having, as a consequence, the same appearance" 
\cite{1auver}. 
In other words, the analogy to the well-known sandpile model suggests
that SOC describes evolution via punctuated avalanches of high mutation rates
over a wide range of sizes, with long intermittent periods of stability, but
without any (long term) time dependence of the avalanche sizes, the
equilibrium periods, or their relative changes, respectively. This overall
large-scale time symmetry is inherent in the model of
Bak and Sneppen, as is easily seen from its results. 

In their one-dimensional case that interests us here, each site on a lattice
representing a species interacts with its two nearest neighbors (thus representing
a food chain, for example). The individual species (sites) are assigned random
numbers between 0 and 1 representing their fitness in the fitness landscape
of the whole evolving pattern of lattice sites. At each time step, the site
with the lowest fitness is chosen to be replaced by a different random number,
thus mimicking Darwinian extinction of the least fit species and its replacement
with another species in the same ecological niche. The same procedure is also
applied to the two nearest neighbors of the least fit species 
to simulate coevolution, so that in sum
the fitness values of three sites are changed per time step. The result of this
procedure is that most species self-organize into uniform distribution above
some critical fitness value, while the few below it produce said punctuated
avalanches of various sizes. The resulting activity pattern for the mutations
clearly shows punctuated equilibrium behavior which is time-symmetric over long
periods of evolutionary time (cf. Fig. 6 in \cite{1bakp}). Moreover, as the
model allows to add together only accumulated mutations at single sites, i.e.,
without differentiating between progress and regression in fitness over time,
the accumulated activity pattern is also time-invariant (cf. Fig. 7 in \cite
{1bakp}), and has recently been identified with the fractal pattern of a
randomized devil's staircase function \cite{1baks3}.

In contrast, we design our toy model in such a way that four major problems
are directly adressed to overcome some weak points in the existing models of
punctuated equilibrium: 
1) To potentially allow for 
the emergence of ever higher complexity, the implicit
overall time-symmetry of the evolving patterns in state-space must be discarded.
2) Accordingly, a plot of evolutionary changes for a single species over time must
not only add accumulated changes (producing a devil's staircase function), but
must be able to differentiate between progress and regression with regard to
some measure of fitness or complexity (cf. Fig. 1 in \cite{1ele} and Fig. 1A in
\cite{snepp}).
3) The strict coupling between mutation and selection, i.e. the ongoing and
strictly time-ordered interplay of both mechanisms, which is often considered
as a necessary prerequisite for evolution, has to be
relaxed to allow for the self-organization of evolutionary patterns.
In this way, coevolutionary processes can lead to an effective
decoupling of mutation and selection.
4) As the model to be designed should depend even less on the concrete content
of micro-evolution, it should be as abstract as possible to study potential
trends in evolution with a high chance for universality, i.e., irrespective
of the underlying detailed processes. We thus aim at a meta-macroevolutionary
model in the sense that we disregard questions on the concrete evolutionary
substrate, thereby concentrating on relative changes with respect to previous
evolutions. In other words, we are interested in simulating the ``evolution of
evolution". Also, as far as the assumption 
holds that the same evolutionary mechanisms
are at work at the levels of populations and species, respectively, our model
for the species level will apply to the population level as well.

To implement the four points just mentioned, we propose the following simple
toy-model for the evolution of evolution.

%********************************************************************2
\section{Hierarchically emergent fractal evolution (HEFE) with noise}

\subsection{Fractal evolution}

We consider the evolution in time of a one-dimensional array of cells
$\rho_0(t,j)$ with $j = 1, \dots,N$ 
representing the number of species
on a coupled map lattice (CML). The nearest neighbor
interaction implemented is of the form
\beq \label{1:origrule} \begin{array}{c}
\rho_0(t,j)=\frac{1}{{\cal N}_0 (t)} 
\{\rho_0(t-1,j) + \de \cdot [\rho_0(t-1,j-1)+\rho_0(t-1,j+1) ] \}\\
j = 1, \dots, N, \qquad \rho_0,\de \in {\cal R}
\end{array} \eeq
with $\de$ as coupling constant and ${\cal N}_0$ 
as normalization factor yielding
\beq \label{1:rho} 
\sum_{j=1}^N \rho_0(t,j) = 1 \qquad \makebox{for all} \,\, t\,.
\eeq

To constrain the values of the sites within a realistic domain,
we introduce a system's {\em internal} 
precision limit $\xi_{max} = 10^{-6}$ with the
effect that each site $\rho_0(t,j) \le \xi_{max}$ is replaced by a white
noise term, generated by $\xi_{max} \cdot RAN$. The random number $RAN$ is
uniformly distributed on the interval $[0,1]$.

In this way, $\rho_0(t)$ assumes values between zero and one representing the
individual species' fitness at time step t relative to the others in such a manner that the
total fitness for the whole array is given by a constant which, without loss
of generality, is defined to be equal to unity.
In other words, just as in \cite{1baks2}, we assume local interaction between
the nearest neighbored species to affect relative fitness. However, as opposed
to \cite{1baks2}, we are not interested in implementing a very specific and
thus extremely model-dependent mechanism of eliminating least fit species, or
the like. Rather, we intend to study the {\em relative frequency of deviations
from some average evolution}, the details of which are not supposed to matter.
To do so, we need to have an evolution of the form of Eq. (1) upon which
certain constraints can be imposed, such that some kind of statistically
controllable but individually contingent deviation can be implemented.

With regard to this, a phenomenon is at hand which has been discovered and
studied by us in some detail, called {\em fractal evolution} \cite{2fg,2fgs1}.
It is a universal dynamic property shown for one-dimensional CMLs with (i) one
or more temporal feedback operations (involving some
memory of the system's states) boosting certain sites to values far above the
average ones, and (ii) a normalization procedure after each time step
as given by Eq. (2). Fractal evolution is characterized by a power-law behavior
of the system's order parameter with regard to a resolution-like parameter
which controls the deviation from an undisturbed (i.e., feedback-less) system's
evolution and provides a dynamically invariant measure for the emerging
spatiotemporal patterns \cite{2fg}. It must be strictly distinguished from
the phenomenon of fractal growth, which describes the accumulation of
micro-patterns into a static structure with (self-similar or other scale invariant)
fractal features. In contrast, fractal evolution is characterized by the fact
that the dynamics of the evolution itself is scale invariant: instead of
generating a ``frozen" fractal object, the generation mechanism itself
is ``fractal", i.e. the chosen resolution itself finally generates the
observed fractal properties. In reference \cite{2fgs1} we showed that
the phenomenon is very robust with respect to variation of the system's variables
and holds also if, instead of a temporal feedback condition, the deviations
from the average evolution are implemented by randomly generating local boosts
to high values for individual sites. In these cases, the only difference to
the feedback situation is a different resulting fractal evolution exponent in
the power law. This means that no matter how the detailed micro-evolutionary
substrate is implemented, i.e. by either entirely deterministic temporal
feedback loops with a memory, or by completely random boosts to produce patterns
deviating from the average development of the system, fractal evolution will
be observed in most cases \cite{2fgs1}.

In the following, we shall choose temporal feedback loops in our model because
they introduce a quality of self-reference for the whole system which can be
expanded upon later (cf. Sec. 2.2). As a concrete motivation for 
doing so, we consider a certain
periodicity of constraining conditions for the whole evolutionary landscape.
Again, the details are not supposed to matter, i.e., the periodicity may refer to
relatively short-term cycles of constraints enforcing functional couplings
within individual species, or externally, in the ecology
or the like. However, the period of constraints
should also not be too short, because otherwise internal fluctuations
would merely produce spurious effects of the internal systems dynamics which 
are of no particular interest here.

In a ``normal" evolution of our landscape 
that were unaffected by such cycles, the
relative fitness values would gradually become adjusted to each other, thus
gradually reaching an ever more smeared-out pattern representing stasis or
equilibrium. However, if a species' fitness value after such a cycle happens to
be the same again as before it, within a 
certain degree of accuracy $1/\ep$, then, even
if the relative fitness value is low, some reproductive isolation must have
been achieved (for example, via economic functional couplings)
that amounts to an acquisition of a relative survival rate far
above the average. In fact, ``molecular cognition" within genomic dynamics
\cite{schust}
already suggests a certain cognitive ability for
each individual site in that it can monitor its own ``fitness" within that of its environment with
a resolution $1/\ep$.
% (For reasons to be elaborated below, we call this a ``first order resolution".)
Maintaining constancy in fitness space within cycle
periods with a constraining and eventually destructive potential, while most of the environment
produces decreasing fitness values, means that an ecological niche has been found
whose fitness potential can be consumed. (The overall relative fitness is thus
decoupled from the individual and momentary survival rate: the former is a
purely coevolutionary measure, while 
the latter refers to a concrete situation in the
larger context of cyclically changing 
constraining conditions of the whole evolutionary
landscape.) In these cases, then, the species' site is a particularly favorable
niche and therefore
enhanced to some boost value, thus providing a fragmented
overall pattern emerging in fitness space
with particular elevated zones of highest survival rates (and,
consequentially, also of fitness).

These zones represent a high possibility for reproductive isolation (for
example on the population level) and thus
also for protected change. Although in principle it can occur at any time or site,
a species' possible mutation, then, is most likely
to accumulate into an evolutionarily conserved change 
(for example a morphological one) if
occuring in such boosted niches of survival rates (and thus also new fitness
values) far above the average. Naturally, if the accuracy $1/\ep$ within which
present and past relative fitness values are observed (i.e., to produce
constancy with respect to the steadily evolving environment) is rather low,
there will be a relatively high chance for various boosts to occur within short
periods of time. If, on the other hand, $1/\ep$ becomes higher 
and higher, i.e., if
the cognitive requirements regarding the resolution become ever 
more demanding, then the chance for boosts
gradually diminishes, and the actually boosted patterns will survive
for longer periods of time.

To implement said temporal feedback operation in our model of Eq. (1), we
install an enhancement (of strength $A_{amp}$) 
of each site $\rho(t,j) > \xi_{max}$ fulfilling an $\ep$-condition

\beq \label{2:full} \begin{array}{c}
\rho(t,j)=\frac{1}{{\cal N} (t)} 
\{ [A_{amp} - \rho_{0,\xi}(t,j) ] \cdot \theta_{\ep} \cdot 
\theta_{mem} \cdot \theta_{\xi} +
\rho_{0,\xi}(t,j) \} \\
\mbox{with}  \\
\rho_{0,\xi} (t,j) = [ \rho_0(t,j) - \xi ] \cdot \theta_{\xi} + \xi \,\,,
\,\, \rho_0(t,j) \,\, \mbox{defined in Eq. (\ref{1:origrule})}\,\,,\\
\xi = \xi_{max} \cdot RAN,\,\, RAN \in [0,1], \,\,
\theta_{\xi} = \Theta(\rho_0(t,j) - \xi_{max})\,\,,\\
\theta_{\ep} = \Theta \left( \ep - 
\left| \frac{\rho_{0,\xi}(t,j)}{\rho(t-t_{mem},j)} - 1 \right| \right)\,\,,
\quad 0 \le \ep \le 1 \, (= 100\%)\,,\quad 
\theta_{mem} = \Theta(t-t_{mem})\,\,, \\
\mbox{and} \quad \Theta(x) = \left\{ \begin{array}{r@{\quad:\quad}l}
1 & x > 0 \\ 0 & x \leq 0 \,\,. \end{array} \right.
\end{array} \eeq
As usual, we impose periodic boundary conditions. Also, we introduce a time span
$t_{mem}$ during which the system's memory is effective. (For example, the latter
may reflect the maintainance of a particular molecular state during one cycle of
the periodical constraining conditions.)

In other words, the enhancement for any site
$\rho_{0,\xi}(t,j) > \xi_{max}$ 
takes place if $t > t_{mem}$ (i.e., the time span during which a system's
memory is effective) and if the condition $\rho(t-t_{mem},j)\cdot (1-\ep) < 
\rho_{0,\xi}(t,j) < \rho(t-t_{mem},j)\cdot (1+\ep)$ is fulfilled.
Note that this kind of feedback operation simulates a selection
process enhancing the values and, consequently, the further evolutions
of those sites which are comparable in fitness with their 
corresponding values in the past
since they fulfill the $\ep$-condition.

The whole normalized system is constrained to the range
$0 \le \rho(t,j) \le 1$. 
Values of $\rho(t,j) \le \xi_{max}$ are, just for simplicity, 
replaced by white noise terms, i.e.,
in that case, whatever kind of noise existing in our system exceeds
a signal generated by the system's rules.
Moreover, a further parameter is introduced as
an {\em external} representation threshold $L >> \xi_{max}$: with it a
background is defined against
which the arising patterns obeying $\rho(t,j) \ge L$
can easily be discerned. Otherwise, e.g. for  $L =  \xi_{max}$,
the evolving systems' pattern would appear as one connected stream and 
could not be discerned into the separated fragments of the landscape's
peaks needed for quantitative analysis.

For the further analysis and without loss of
generality (cf. the discussion below) 
we choose a default set of parameter values:
\beq \begin{array}{l} \label{2:param}  
\mbox{Number of sites (dimension)} \qquad N = 120  \\
\mbox{Initial values (before normalization)} \quad 
\rho(0,40)=0.1\,\,, \rho(0,60)=0.9\,\,, \rho(0,100)=0.3 \\
\mbox{Mixing parameter (coupling constant)} \qquad
\delta = 0.03 \\
\makebox{Boost value (enhancement)} \qquad A_{amp} = 100  \\
\mbox{Size of memory} \qquad t_{mem} = 200 \\
\mbox{Threshold value} \qquad L = 10^{-2} \\
\end{array} \eeq

\begin{figure}
\hspace{-.7cm}
\epsfig{file=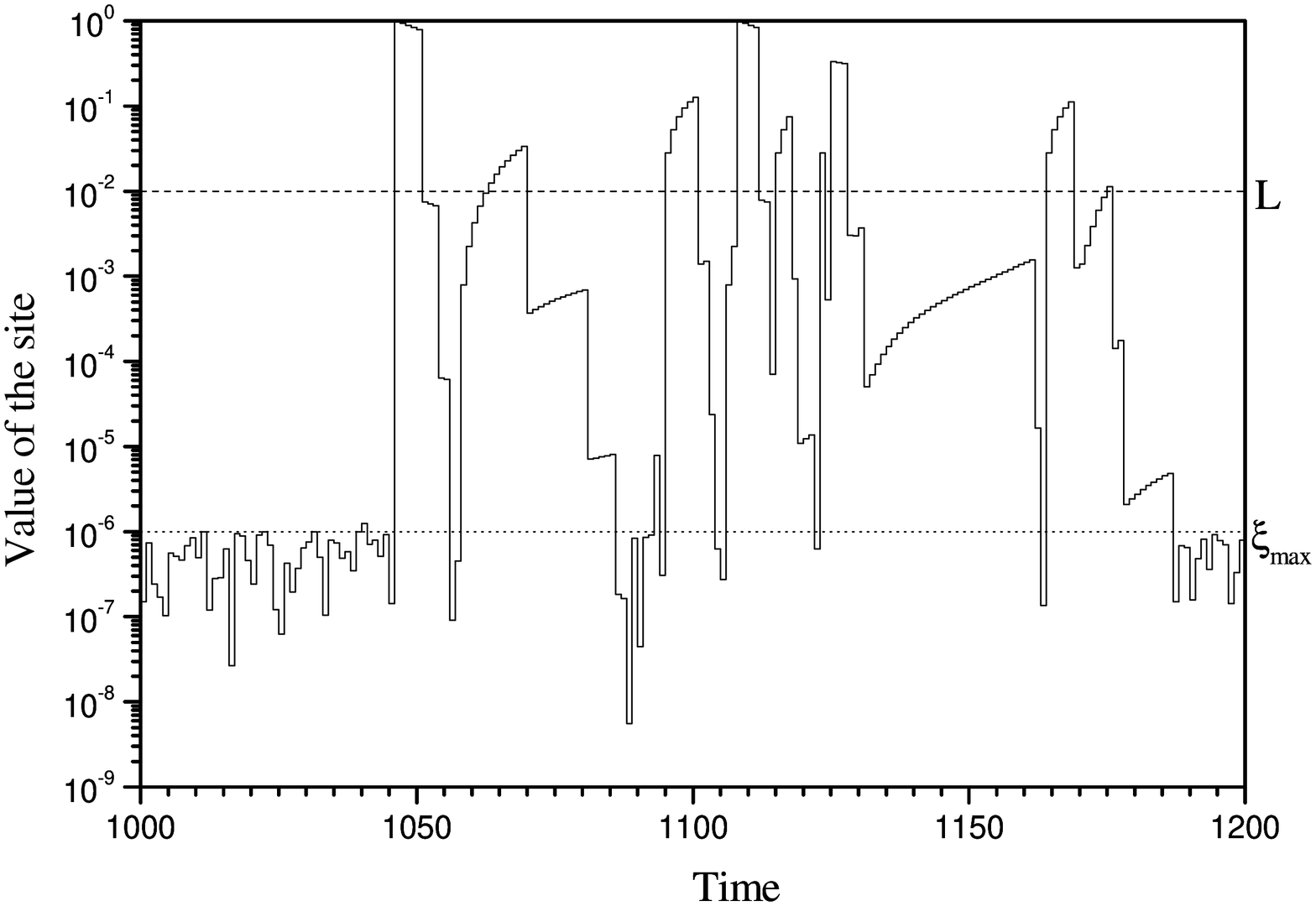,height=14.cm, width=11.cm,angle=-90}
\vspace{-1.cm}
\caption{
Temporal evolution of an arbitrary site ($\sharp 48$), initiating from
the noise domain below the threshold $\xi_{max}$, and later entering the 
region of
an observable pattern above the representation threshold $L$.}
  \label{fig:1}
\end{figure}

In Fig. (\ref{fig:1}) the time evolution due to Eq. (\ref{2:full}) of one arbitrarily
chosen site (i.e., site $\sharp 48$ in our case)
is displayed together with the noise threshold (dotted line) and the
external representation threshold used in the further graphical representation of
the plots (dashed line). Due to Eq. (\ref{2:full}), 
at time steps $t = 1046$ and $t = 1108$ the site
is boosted up to a value of order $O(1)$ after normalization, 
whereupon the typical
decrease due to the diffusion rule (\ref{1:origrule}) is observed until
the boost of some other site within the array suppresses it by the factor
$A_{amp} = 100$ due to the normalization procedure. 
The feature of repeated staircase-like increases of the site's values
comes from contributions of
neighbored large-valued sites.

\begin{figure}
 \centering
\hspace{5.cm}
\epsfig{file=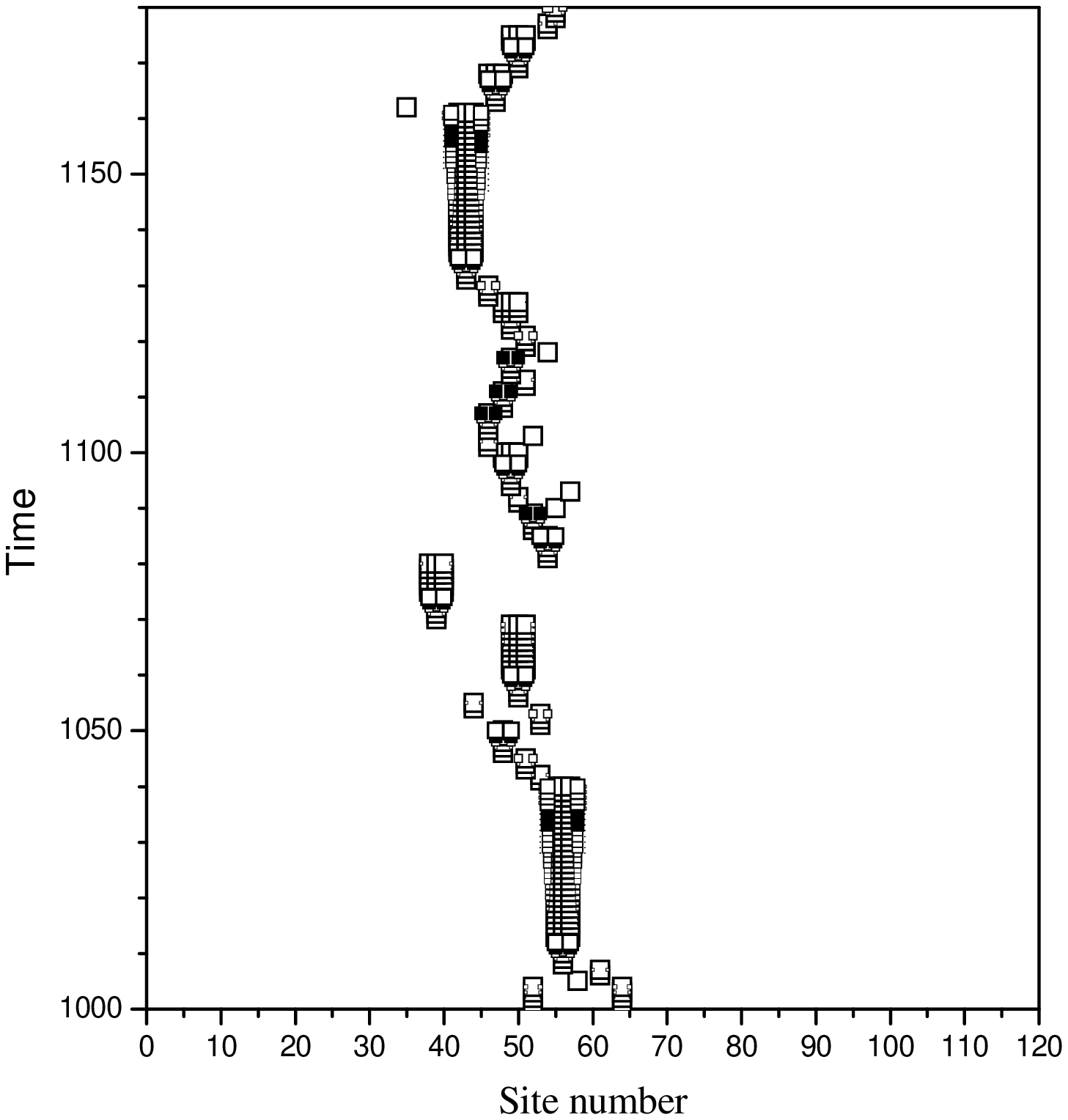,height=15.cm, width=11.cm,angle=-90}
\vspace{-1.cm}
\caption{Density plot of the evolving system accentuating the range of the
box values from $L = 10^{-2}$ to $10^{-1}$. Values larger than 
$10^{-1}$ are displayed with the maximal box size. The relative interval
width $\ep$ is chosen as $\ep = 30\%$.}
  \label{fig:2}
\end{figure}

A spatiotemporal plot of all sites above the threshold value $L$
is shown in Fig. (\ref{fig:2}). Contrary to the signature of fractal evolution
introduced in \cite{2fg,2fgs1}, the fragments gather around a narrow
array and are not distributed over the whole spatiotemporal plane. This effect
represents a certain ``taming" of evolutionary chance and
comes from the restriction requiring that all sites be larger than the noise
threshold $\xi_{max}$ for their possible enhancement. (As stated above, eventual
values below the noise threshold thus contribute only to the noise itself,
but cannot be boosted to values dominating the whole evolution.)
Since the range of those values
which are checked by the $\ep$-condition of (\ref{2:full}) contains only
six orders of magnitude, a boosted site will be placed in the spatial
neighborhood of a fragment $t_{mem}$ time steps ago. As will be shown
below, this ``stream-like'' feature in the fitness landscape
is crucial for the punctuated equilibrium 
behavior during the systems evolution.

\begin{figure}
% \centering
\epsfig{file=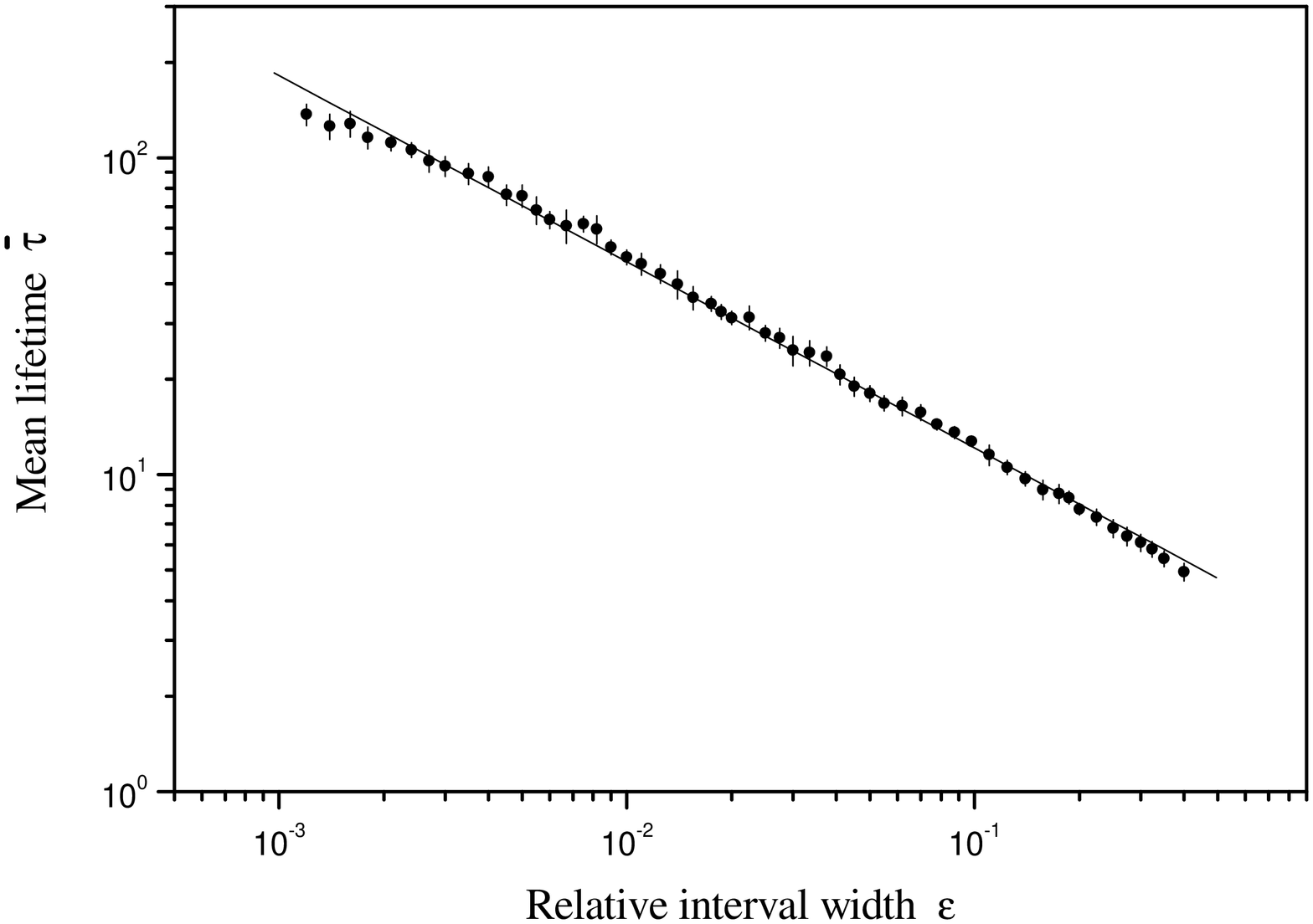,height=14.cm, width=11.cm,angle=-90}
%\vspace{-1.cm}
\caption{Log-log plot of the mean lifetimes $\tilde{\tau}$ of the fragments
versus the relative interval width $\ep$. System parameters are taken
from Eq. (\ref{2:param}). The fractal evolution exponent is estimated
by the slope of the fit curve.}
  \label{fig:3}
\end{figure}

In Fig. (\ref{fig:3}) the system's global property of fractal evolution is shown. %, i.e.,
The mean lifetime $\tilde{\tau}$ of the patterns which is estimated
by the arithmetic mean of the maximal temporal extension of all fragments 
for a chosen
$\ep$ (cf. Fig. (\ref{fig:2})) scales according to

\beq \label{2:power}
\tilde{\tau} = a \cdot \ep^b, \qquad \mbox{with} \quad a = 2.69 \pm 0.05, \,\,
b = -0.620 \pm 0.005 \,\,,
\eeq
practically irrespective of variations of the systems variables or the
initial conditions. The power-law behavior of the system's order
parameter $\tilde{\tau}$ with regard to the resolution-like parameter
$\ep$ 
%indicates a self-organized critical state from the viewpoint of
%the SOC-theory \cite{1baks1} which is in fact reached
provides a dynamically invariant 
measure for the emerging spatiotemporal patterns given by the fractal
evolution exponent $b$. As will be discussed in Sec. 2.2,
fractal evolution in its dynamic sense
is obtained only if the restriction to one specific value of $\ep$
is abolished.
Note that the system evolving under a chosen
$\ep$ does not evolve automatically towards a critical state in the sense
of the SOC theory \cite{1baks1} since the avalanches of the fragments' sizes
do not occur in our system on all scales.

\begin{figure}
% \centering
%\hspace{1.cm}
\epsfig{file=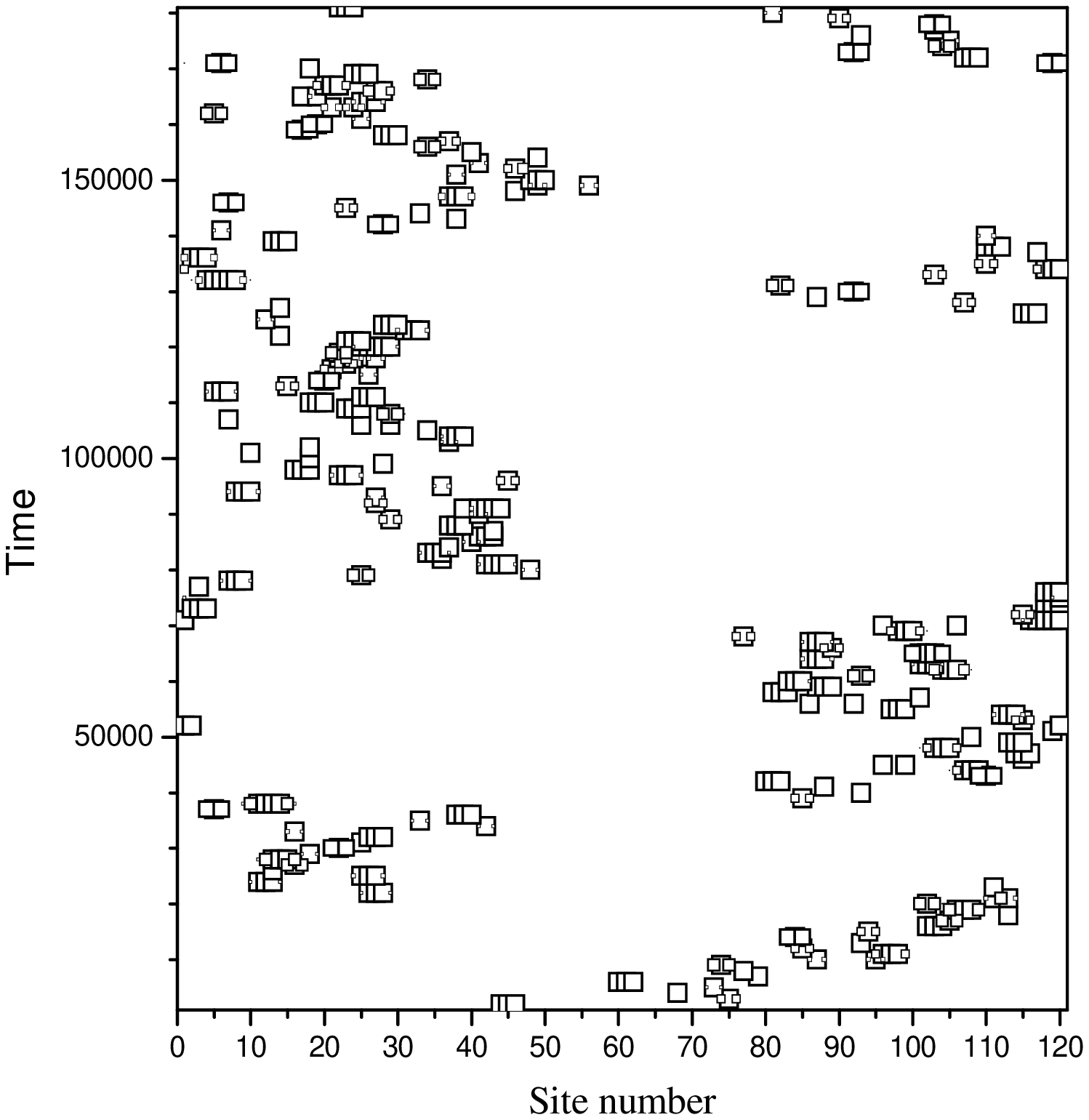,height=15.cm, width=11.cm,angle=-90}
\vspace{-1.cm}
\caption{Same as in Fig. (\ref{fig:2}), but now only for each 1000th time step, in order to see
the long-term evolution. The narrow band of fragments in Fig. (\ref{fig:2}) eventually
enters practically all regions of the array in a random-walk-like manner.}
  \label{fig:4}
\end{figure}

Although the two concurring conditions in Eq. (\ref{2:full}), i.e. the
temporal $\ep$-condition and the noise-threshold condition, constrain the
evolving patterns to lie within a narrow range, the long-term 
behavior as shown in Fig. (\ref{fig:4})
exhibits a random-walk-like movement of the stream of fragments
across the whole array. The invariance of this ergodic-like 
property under variation of the systems parameters has been
checked empirically. Even the replacement of white noise values for sites
below $\xi_{max}$ by a constant value (by $10^{-7}$, for example),
provides the observed ergodicity, although the stream of fragments then evolves
symmetrically due to the totally symmetric evolution rule and the uniform
 background. Therefore, in the long run, practically each site/species
gets the chance to obtain a high survival rate and, consequently,
to evolve in a high fitness potential.

\begin{figure}
 %\centering
\hspace{-1.cm}
\epsfig{file=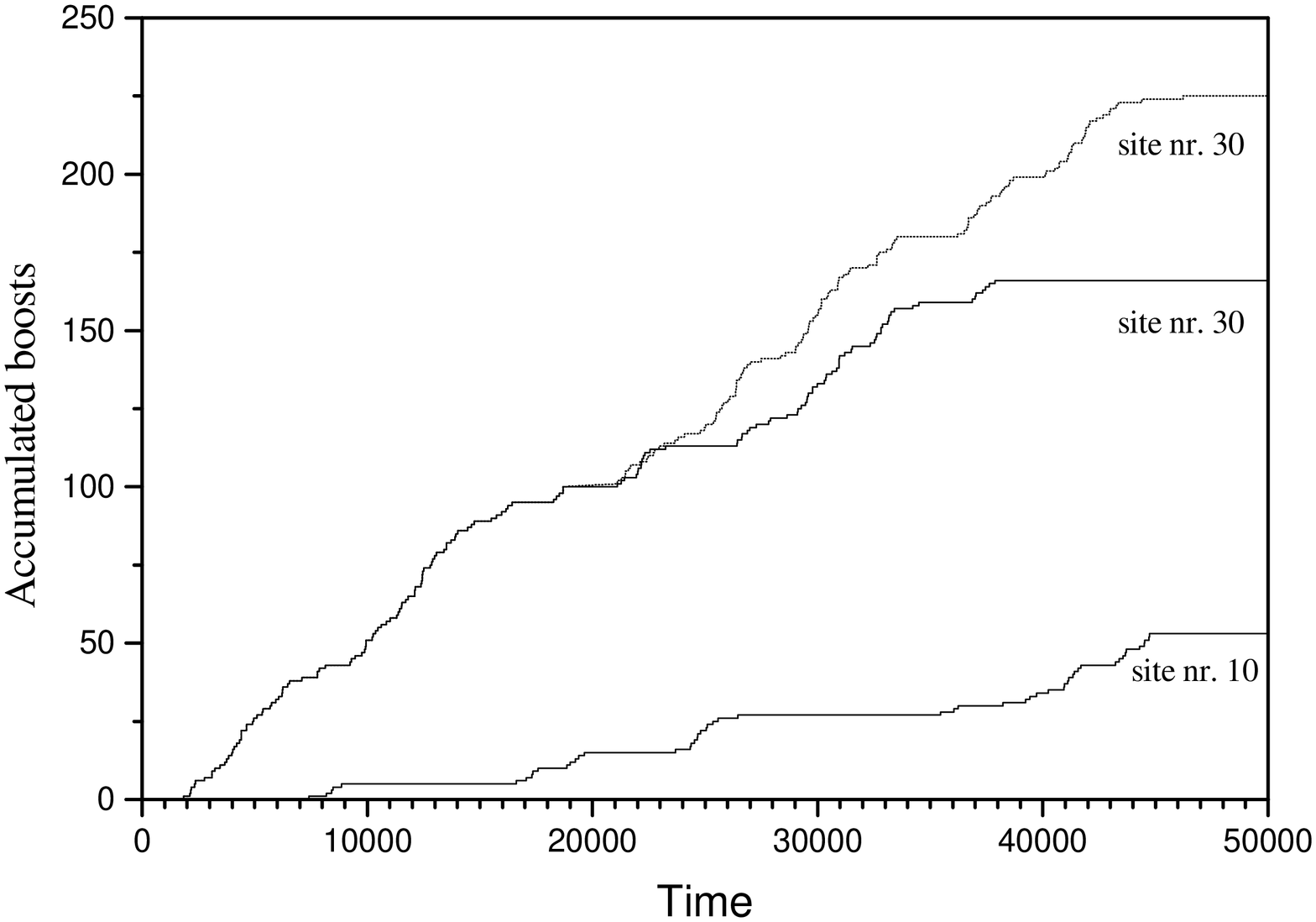,height=15.cm, width=11.cm,angle=-90}
\vspace{-.5cm}
\caption{The accumulated boosts for two different sites show punctuated equilibrium
behavior, i.e., periods of stasis are interrupted by high
activity of boosts including eventual mutations. 
The dashed line shows the effect of omitting the 100th boost for
site $\sharp 30$.}
  \label{fig:5}
\end{figure}

Any enhancement of a site represents a strong
intervention for the system as a whole (due to the normalization) and for the 
represented species itself, of course, since it can be interpreted both as a
drastic change in coevolutionary fitness or as a site-bound beneficial mutation, 
respectively. Consequently, the time span between
two succeeding boosts for any site offers both the opportunity for further
beneficial selection during its evolution, or towards ever decreasing fitness,
and ultimately being swept below the noise limit (``detrimental extinction"). 
In Fig. (\ref{fig:5}) the accumulated
boosts are displayed for a time interval of 50000 time steps.
Similar to the SOC models \cite{1bakp,1baks3}, each site exhibits a punctuated
equilibrium behavior with respect to the time plateaus indicating the time
difference between two drastic evolutionary activities. Due to the chosen
intial conditions and within the first 50000 time steps, 
site $\sharp 30$ happens to experience many more changes than
site $\sharp 10$.
Since the stream of fragments and thus the main boost activities 
will reach also 
previously quiescent regions after some time (cf. Fig. (\ref{fig:4})),
the accumulated boosts of site $\sharp 10$ will eventually catch up with 
those of site $\sharp 30$. The dashed curve in Fig. (\ref{fig:5}) shows the effect
of omitting only one boost (i.e., the 100th one in our case) during the site's 
evolution. The observed dramatic change of the accumulation curve
demonstrates just one example of the crucial role of contingency in the 
systems dynamics. 
\begin{figure}
% \centering
\hspace{-1.cm}
\epsfig{file=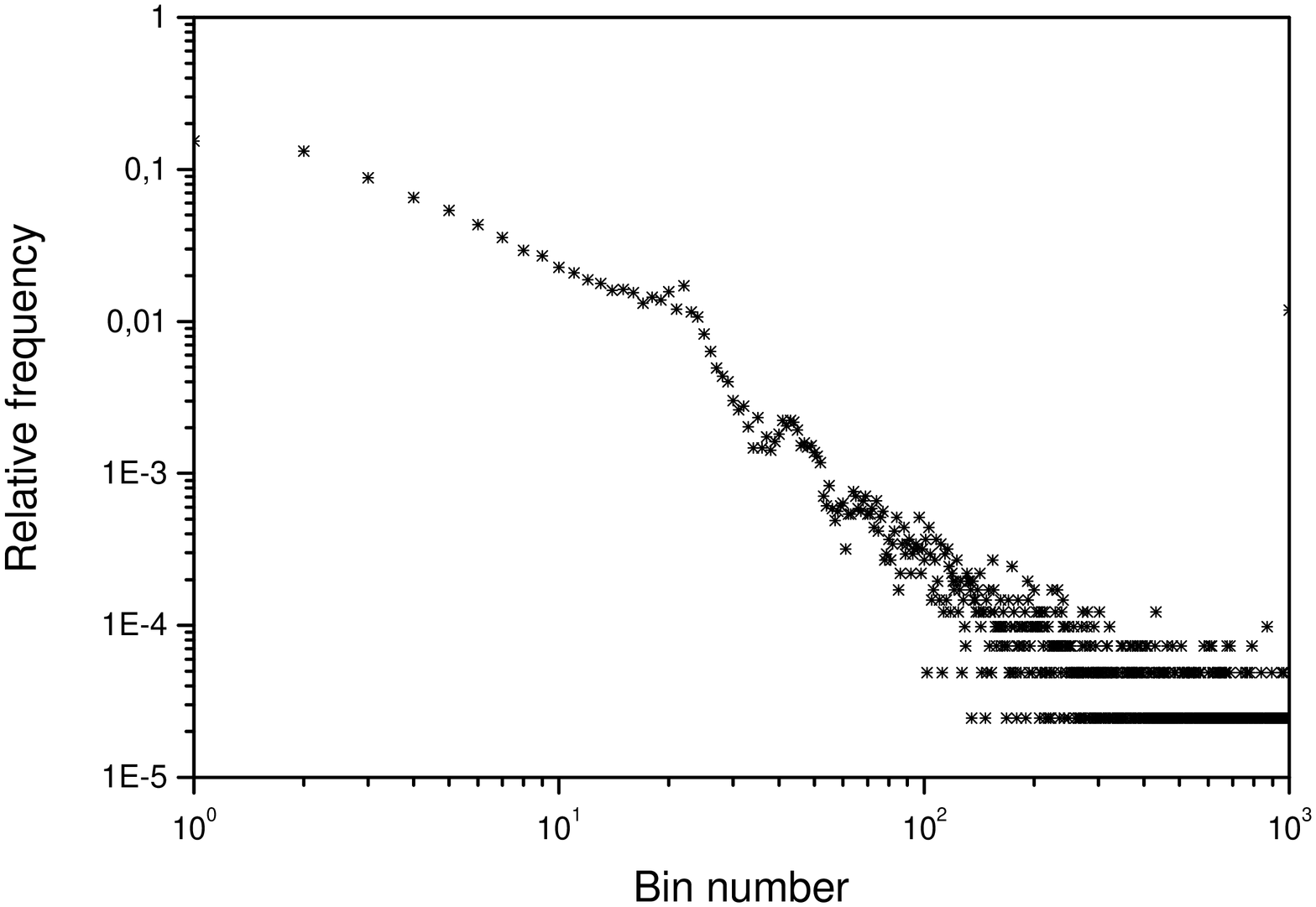,height=15.cm, width=11.cm,angle=-90}
\vspace{-.5cm}
\caption{Relative frequency distribution of the sizes of the time plateaus in Fig. 
(\ref{fig:5})
according to their size, represented by bin numbers with a bin width of 10.
Bin $\sharp 10$, e.g., contains all sizes from 90 to 99. Bin $\sharp 1000$
contains the sizes 9990 to 9999 and, additionally, all other ones larger than
9999. The data were taken over a period of $2.10^7$ time steps.
Note the appearance of the first peak around bin $\sharp 20$ which is
explained in the text. The left hand side
of the figure represents behavior according to a randomized devil's staircase
function.}
  \label{fig:6}
\end{figure}
The analysis of the distributions of time plateaus for an arbitrarily chosen
site yields a (discretized) shape as shown in Fig. (\ref{fig:6}). Each bin contains
the relative frequency of ten succeeding plateau sizes. Apparently,
practically all sizes of periods of stasis are present, the data for bin
$\sharp 2$ up to about $\sharp 20$ even lying on a staight line in a
log-log plot. The peaks around bin $\sharp 20$, i.e. for plateau
sizes of order $O(200)$, and another, weaker one
around bin $\sharp 40$ (plateau sizes of order
$O(400)$) represent a direct effect of the chosen size of the system's memory
$t_{mem} = 200$. Succeeding boosts occurring with temporal distances
of order $O(k \cdot t_{mem})$, $k = 1,2, \dots$ are causally related
due to the evolution rule (\ref{2:full}) and the values of the systems
variables. The remaining boost distances are clearly not directly related
and their distribution represents a genuine emergent property of
the systems dynamics. 
\begin{figure}
% \centering
\hspace{-1.cm}
\epsfig{file=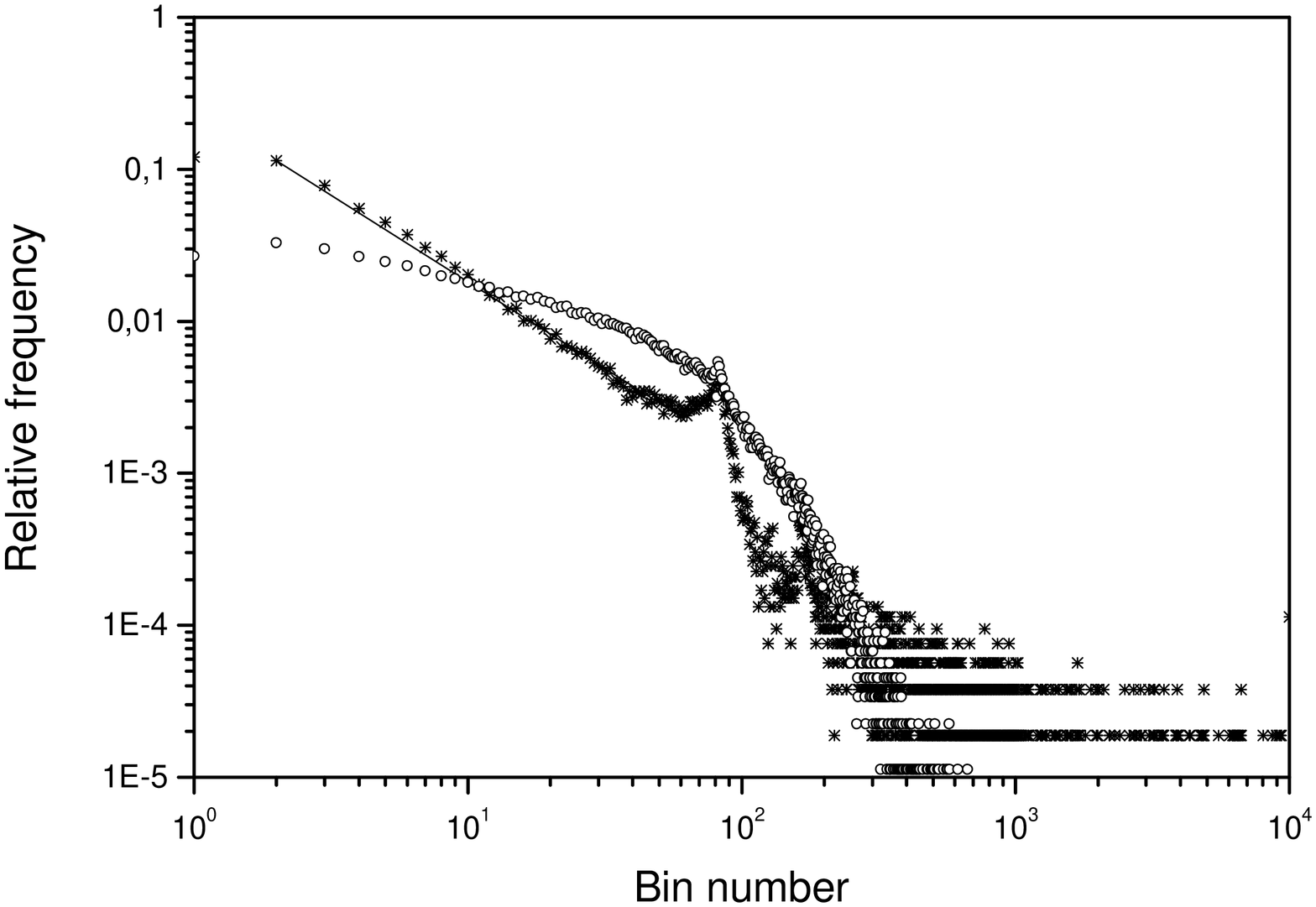,height=15.cm, width=11.cm,angle=-90}
\vspace{-.5cm}
\caption{Same as in Fig. (\ref{fig:6}), but for a larger memory $t_{mem} = 800$ and for 
a larger number of bins to make the occurrance of very large time plateaus
more clear. The first peak now shows up at around bin $\sharp 80$. 
The circled data represent the originally introduced noise-free
model of fractal evolution \cite{2fg,2fgs1} indicating a sharp cutoff of the
plateau sizes. The fit indicates the slope due to a randomized devil's
staircase function.}
  \label{fig:7}
\end{figure}
As long as we use a systems memory, this perturbation 
of causally unrelated boost distances by causally related ones seems to
be inevitable, but the moment of the peak's first occurrence
can be shifted towards larger plateau sizes as shown in Fig. (\ref{fig:7}). In that case,
a memory depth of $t_{mem} = 800$ was chosen. Accordingly, the peak
shows up at about bin $\sharp 80$. Up to this value, the frequency $f(s)$ of
plateau sizes $s$ scales according to 
\beq
f(s) \propto s^{-1.17 \pm 0.01}
\eeq
thus indicating a randomized devil's 
staircase function \cite{mandel,1baks3}. The exponent obtained
above represents a robust property of the system in spite of its inherent
contingency indicated by Fig. (\ref{fig:5}).
For comparison, the circled data in Fig. (\ref{fig:7}) represent the
distribution of boost distances due to our evolution rules used earlier
\cite{2fg,2fgs1}, where no noise threshold condition had been
incorporated. The observed
fragments' distribution all over the spatiotemporal plane in the noise-free
model is reflected in the sharp cutoff of frequencies for plateau sizes beyond
$s \approx 10^4$. 

%\newpage
\subsection{Self-organized evolution of fractal evolution}
As has already been remarked in the 
previous section, fractal evolution in the true
sense of the word is effectively realized only if the interval width $\ep$ is
allowed to acquire different values during the systems evolution.
According to a recently introduced idea \cite{2fgs2,2fgs3}, to be further
elaborated here, we impose 
an additional long-term control cycle leading to an adaptive change of $\ep$ 
on the next
higher level of the system's hierarchical nesting of feedback loops. 

We thus let the system be driven by fluctuations of 
$\tilde{\tau}_{exp}$ measured with the same $\ep$ 
during two succeeding long time spans 
or generations $(n, n+1)$. The difference
$\tilde{\tau}_{exp}(n+1) - \tilde{\tau}_{exp}(n)$
%, which reflects a difference
%in second order resolution and which emerges from the coevolution of the whole
%fitness landscape, 
is practically
always nonzero due to the finite time interval of data taking. To implement
an adaptive change of $\ep$ or resolution $1/\ep$, respectively,
with regard to an ever changing ruggedness of the whole fitness landscape, this difference
is fed back into
the consecutive value of $\ep$ via the difference quotient of the power-law
(\ref{2:power})
\beq
\frac{\Delta \tilde{\tau}}{\tilde{\tau}} = b \frac{\Delta \ep}{\ep}\,\,,
\eeq
leading to
\beq \label{2:delep}
\Delta \ep = \ep_{n+1} - \ep_n = \frac{\ep_n}{b} \cdot 
\frac{\tilde{\tau}_{n+1} - \tilde{\tau}_n}{\tilde{\tau}_n}\,\,.
\eeq
Since the last term of Eq. (\ref{2:delep})
is generally non-zero due to the fluctuations of the mean lifetime, a different
relative interval width results governing the two succeding generations, and
is determined by
\beq
\ep_{n+2} := \ep_{n+1} = \ep_n \left( 1 + \frac{1}{b} \cdot
\frac{\tilde{\tau}_{n+1} - \tilde{\tau}_n}{\tilde{\tau}_n} \right) \,\,.
\eeq
The geometrical analogy to the analytical determination of the
next value of $\ep_{n+2}$ or $\ep_{n+1}$, respectively, 
is given as follows: the latter lies at the intersection of
the tangent at $(\ep_n,\tilde{\tau}_n)$ belonging to the theoretical
curve $\tilde{\tau}_n = a'\cdot \ep^b_n$ on one hand, and the horizontal line
through $\tilde{\tau}_{n+1}$ on the other. 
%If the data pair
%$(\ep_n,\tilde{\tau}_n)$ does not lie on the curve due to
%Eq. (\ref{2:power})

As a main result, we obtain a progressive alteration of 
$\ep$ or the resolution $1/\ep$, respectively, within the system's evolution which we denote as ``hierarchically
emergent fractal evolution'' (HEFE). In effect, HEFE constitutes
a fluctuation driven and adaptive mechanism for the gradual increase 
of the mean lifetimes of the emerging patterns, 
i.e. for the expansions of regions of high fitness in the fitness landscape.
\begin{figure}
% \centering
\hspace{-1.cm}
\epsfig{file=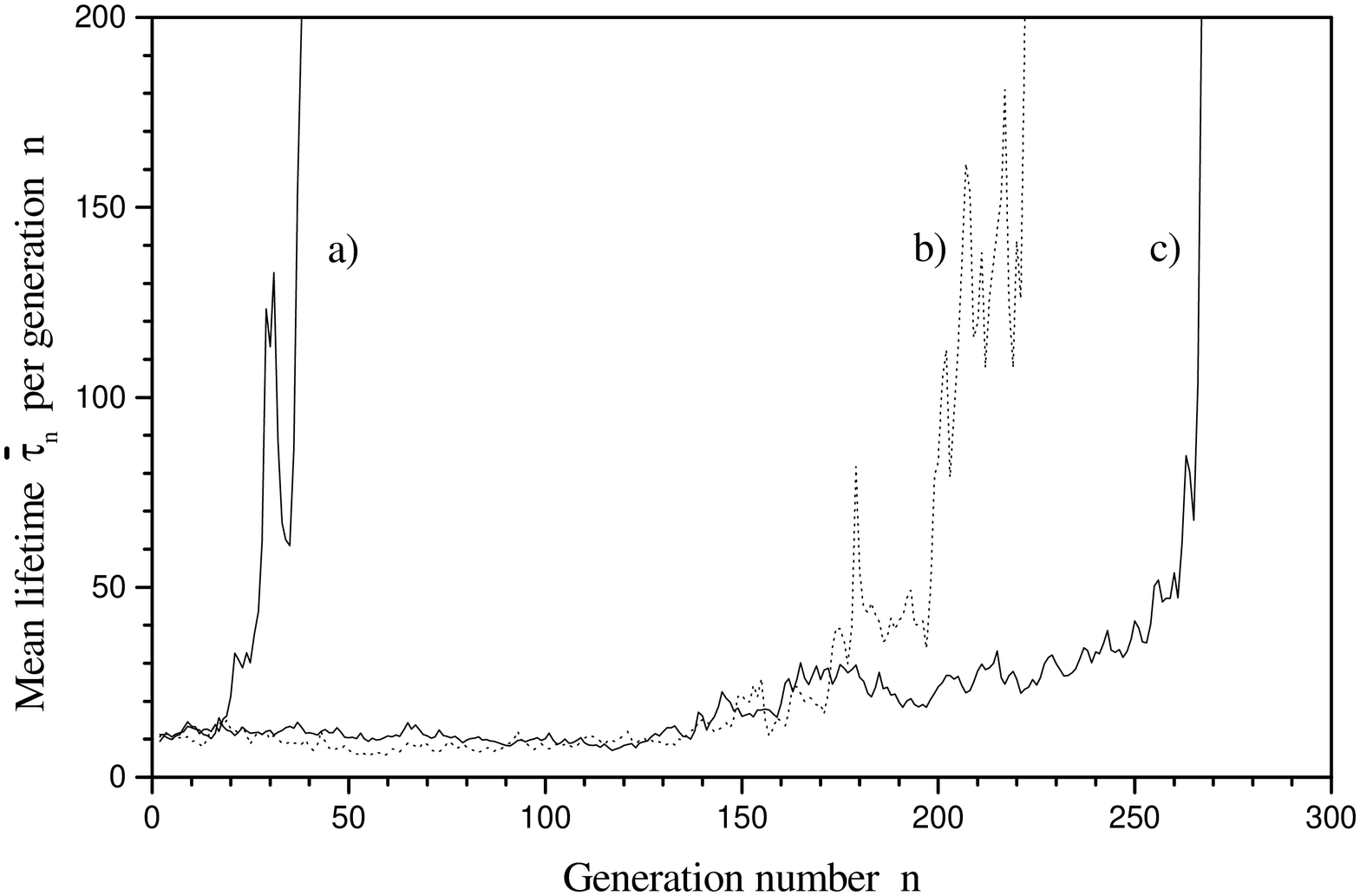,height=15.cm, width=11.cm,angle=-90}
\vspace{-.5cm}
\caption{Hierarchically emergent fractal evolution (HEFE) for three different values of
time spans of data taking with particular $\ep$ (``generations'').
The time span for each generation is given by
2000 time steps for curve a), 5000 for curve b), and 10000 for curve c).
The starting value of $\ep$ for all
curves is $\ep_0 = 10\%$. Apart from the obvious time asymmetry of the evolution,
note also the characteristic stepwise increases as well as decreases of the
mean lifetimes per generation, thus indicating two opposing trends in the
short-term dynamics. In the long run, however, the mean lifetime is bound to
increase.}
  \label{fig:8}
\end{figure}

In Fig. (\ref{fig:8}), three examples of temporal evolutions of the mean lifetime per generation
are shown according to three different time spans of each
generation during which the mean lifetime is determined. Although fluctuations 
are allowed with equal opportunity for decreasing and increasing lifetimes,
respectively, a
long-term increase of the mean lifetimes is observed practically in all
cases and has its origin in the functional property of the power-law
(\ref{2:power}): Denoting the deviation of the mean lifetime
by $\tilde{\tau}_{n+1} - \tilde{\tau}_n =: \eta^{\pm} \cdot \tilde{\tau}_n$,
and assuming one positive-valued deviation $\eta^+$ for the consecutive
generations $(n,n+1)$ and a negative-valued one $\eta^-$ of about equal size 
$\eta^+ \approx -\eta^-$ for the succeeding
ones $(n+2,n+3)$, one obtains altogether a clear tendency towards decreasing relative
interval widths
\beq
\ep_{n+4} \approx \ep_n (1-|\frac{\eta^{\pm}}{b}|^2) < \ep_n \,\,,
\eeq
leading to an average increase of the mean lifetime $\tilde{\tau}$
per generation. If $\tilde{\tau}$ would depend, e.g., exponentially on $\ep$, 
like $\tilde{\tau} \propto e^{-\ep}$, $\ep$ would only oscillate
between two fixed values.

For relatively
small time spans, i.e. for curve a) with 2000 time steps per generation,
the fluctuation of the lifetimes is larger, of course, than for 10000
time steps. 
Therefore, the observed drastic increase
of the mean lifetimes will occur sooner. In sum, the systems dynamics
makes possible a long-term growth of each species' lifetime and thus also of
its complexity: larger lifetimes of high fitness periods inrease, e.g. via
``learning", both the cognitive performances as well as the probabilty for
survival of the species and, consequently, also the probability for genetic improvement
via mutational mechanisms. As far as the cognitive functions are concerned,
their improvement increases the lifetimes of high fitness values, and so on.
We reach the limits of our model when the mean
lifetimes in Fig. (\ref{fig:8}) reach the size of the system's memory $t_{mem}$, because then
fractal evolution gets disturbed \cite{2fgs1} and new boundary conditions would have to be
considered that might constrain the further evolution effectively.

Finally, we briefly discuss the evolution of the system's entropy, which
measures the fitness of the whole ensemble of species, and is simply
chosen as the relative Shannon information entropy \cite{2fgs1}
\beq
\frac{S}{S_{max}} = \frac{-1}{\ln N} \sum_{j=1}^N \rho(t,j) \cdot
\ln \rho(t,j) \,\,.
\eeq

\begin{figure}
% \centering
\hspace{-1.cm}
\epsfig{file=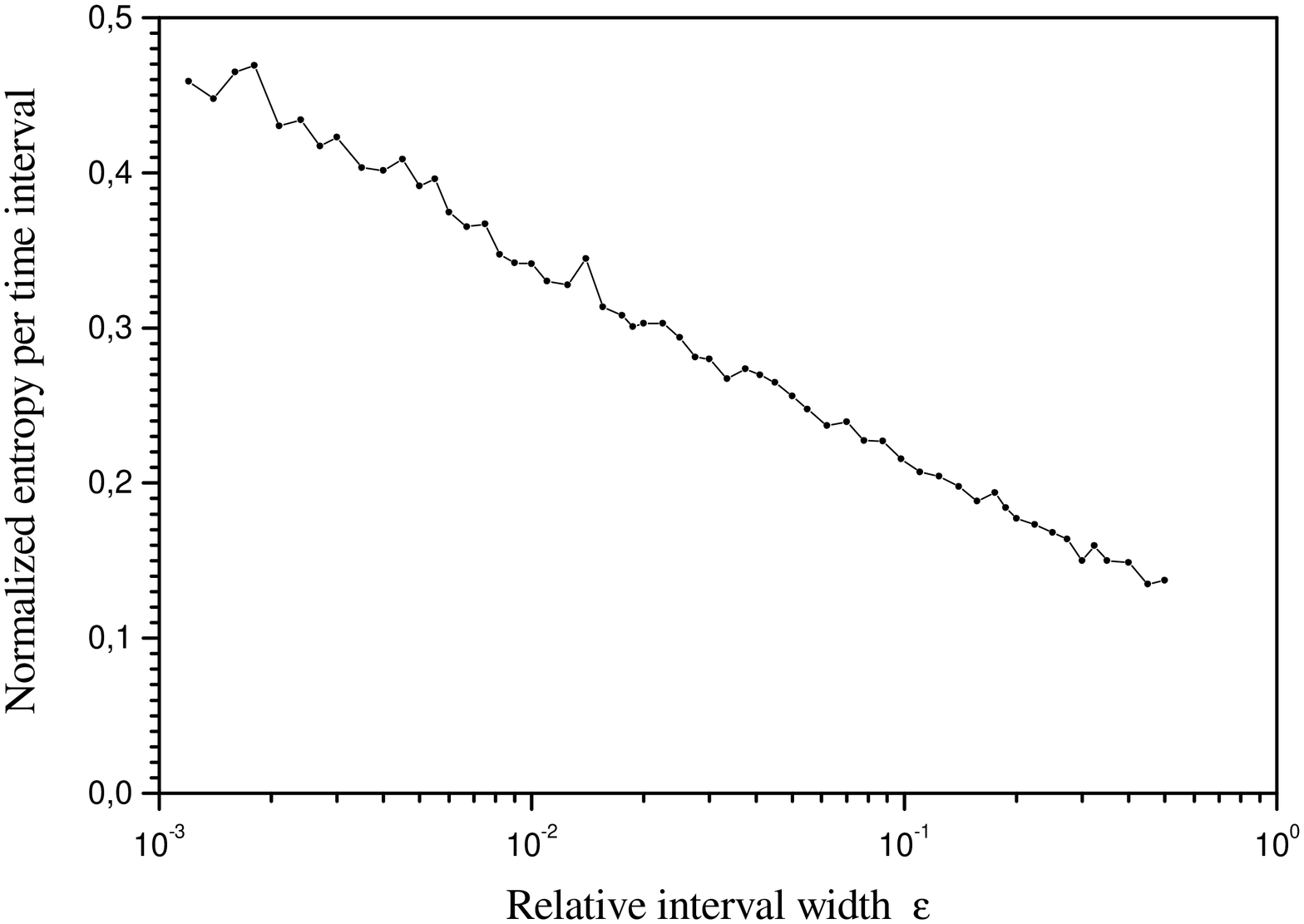,height=15.cm, width=11.cm,angle=-90}
\vspace{-.5cm}
\caption{Normalized entropy per 10000 time steps versus the relative interval width
in a semi-logarithmic plot.
Since the long-term behavior of HEFE develops towards larger mean 
lifetimes $\tilde{\tau}$, and, therefore, 
towards smaller values of $\ep$, the entropy 
of the evolving system increases on the average.}
  \label{fig:9}
\end{figure}

We can allocate the various
activities within the systems evolution with regard to their
specific entropy behavior in the following way. A site's enhancement, i.e.,
an increased single species' fitness value,
lowers the system's entropy at that time step
due to the enlargement of the values' differences within the array.
The effect of the noise threshold and the decay of the fragments due to 
Eq. (\ref{1:origrule}) are entropy increasing processes. Since 
the systems evolution with a given fixed value of $\ep$
yields a stationary pattern production characterized by 
the fractal evolution exponent, an average value of the above described
entropy contributions summed
up for an appropriate long time interval is obtained for that specific
$\ep$.
The larger the value of $\ep$, the shorter the mean lifetime of the fragments
and, consequently, the more rugged the fitness landscape will appear,
thereby lowering the normalized entropy per time interval (cf. Fig. (\ref{fig:9})).
With respect to the HEFE-mechanism, this means that the total entropy
per time interval increases on the average due to the growth of the 
mean lifetimes. That is, within the scenario of macroevolution this effect can
be considered as tendency towards maximal fitness of the whole ensemble of
species under constant boundary conditions. Thus,
{\em internal} evolutions within a species towards higher complexity are
accompanied by a complementary, {\em external} increase of the
entropy of the fitness landscape.

\section{Conclusions}
The scaling behavior of certain observables
in dynamical systems is of particular relevance in a wide range of systems,
be they geophysical, economical or biological, for example. 
Within the theoretical framework of the SOC theory some features of
macroevolution have recently also become testable via a kind of ``experimental
paleontology'' \cite{lens}. They are made plausible by the concept
of a system's critical state which is 
reached after a certain evolutionary time, and which thereby exhibits a
power-law behavior of the frequency of mutation activities with regard to their
sizes. Yet they suffer
from being trapped within the steady-state property of the system's dynamics.

In this paper we have proposed a model for macroevolution based on
concepts differing from the SOC model. Apart from some basic qualitative 
similarities (cf. Figs. (\ref{fig:4}) and (\ref{fig:5})), we obtain substantially
different results, primarily with regard to the
time-invariant behavior of the system's
emerging order parameter: the power-law obtained in our model reflects a
universal feature named ``fractal evolution", which can be based on a variety
of underlying mechanisms, including the option of a memory-based feedback
operation. The four major problems specified in the introduction in order to 
overcome some weak points in existing models of macroevolution are solved
by the implementation of a second feedback operation. The latter transforms the
mere functional property of ``fractal evolution" of the system's fitness 
landscape via a fluctuation-based iterative mechanism into the dynamical process
 called
``hierarchically emergent fractal evolution'' (HEFE). In this way, a self-organized 
potential growth emerges on a long-term scale, such that the internal 
complexity of each species may gradually increase. However, throughout this work we
have deliberately avoided a narrow 
definition of the term ``complexity'', or a choice
among the proliferating definitions in the literature, respectively 
\cite{bates}. 
Instead, we have merely considered beneficial 
fitness domains for each species in the sense of potentials 
to evolve towards higher complexity, irrespective of the
underlying concrete (self-organized) mechanisms. One main result of our study
is that in our model such potentials grow with elapsing time.

We also emphasize the important role of randomness in our model, which is
effective on various hierarchical levels. At the basic level
of pattern generation, we observe a strong sensitivity of the patterns'
signature to slight variations of the initial conditions, similar to
the features of deterministic chaos. An emerging global measure,
irrespective of local system fluctuations,
has been obtained in the form of the fractal evolution exponent. The incorporation of a
noise threshold constraining
the range of operational fitness values for each species
enables the systems' patterns to walk
randomly across the whole array (cf. Figs. (\ref{fig:2}) and (\ref{fig:4})), thereby producing activity/mutation
patterns obeying a randomized devil's staircase function. As in the SOC model
\cite{1bakp,1baks3},
small deviations in the order of mutation events produce significant
deviations from the course of the accumulated activity pattern (cf. Fig. (\ref{fig:5})). The globally
invariant measure in this case is given by the slope of the distribution
function for the frequency of the periods of stasis versus their sizes (cf. Figs.
(\ref{fig:6}) and (\ref{fig:7})). Moreover,
at the level of the self-organized change of the 
resolution-like parameter
$\ep$, the natural fluctuation of the order
parameter $\tilde{\tau}$ within two runs (``generations'') under the 
same conditions iteratively serves as driving force 
towards a changing $\ep$.
The direction of this change appears on small time scales
as a random walk. However, the long-time tendency towards the irreversible
decrease of $\ep$, and thus towards the increase of the mean lifetimes of 
the high fitness domains for each species, respectively, is firmly anchored in
the HEFE mechanism (cf. Fig. (\ref{fig:8})). More generally speaking, it is rooted
in the basic mathematical properties of the power-law describing
fractal evolution, and it leads in the long run to a mutual positive 
enhancement of potential cognitive abilities for all species on one hand, and 
lifetimes with high fitness values on the other.

Finally, the potential development of individual species towards gradually
higher complexity (as represented by an evolution towards ever lower values of
$\ep$) is accompanied by a tendency towards highest entropy values for the whole
fitness landscape under constant boundary conditions (cf. Fig. (\ref{fig:9})). The HEFE
mechanism can thus be interpreted such that a single species is constantly
engaged, via the possibilities of beneficial selection versus detrimental 
extinction, in a struggle for higher fitness values, whereas there exists a
global tendency of ever higher fitness values to emerge for the whole
ensemble of species.

%************************************************* *****References


\begin{thebibliography}{99}
%
\bibitem{1auver}
Ausloos M. and Vandewalle N., Algorithmic models for evolution,
{\em Europhys. News} {26} (1995) pp. 55-56 
%
\bibitem{1bakp}
Bak P. and Paczuski M., Complexity, contingency, and criticality,
{\em Proc. Natl. Acad. Sci.} {\bf 92} (1995) pp. 6689-6696 
%
\bibitem{1baks2}
Bak P. and Sneppen K., Punctuated equilibrium and criticality in a 
simple model of evolution, {\em Phys. Rev. Lett.} {\bf 71} (1993) 
pp. 4083-4086 
%
\bibitem{1baks1}
Bak P., Tang C. and Wiesenfeld K., 
Self-organized criticality: an explanation of $1/f$ noise, 
{\em Phys. Rev. Lett.} {\bf 59} (1987) pp. 381-384, 
Self-organized criticality, {\em Phys. Rev.} {\bf A38} (1988) pp. 364-374.
%
\bibitem{bates}
See, e.g., Bates J.E. and Shepard H.K., Measuring complexity using information
fluctuation, {\em Phys. Lett.} {\bf A172} (1994) pp. 416-425. 
%
\bibitem{1baks3}
Boettcher S. and Paczuski M., Exact results for spatiotemporal correlations
in a self-organized critical model of punctuated equilibrium, 
{\em Phys. Rev. Lett.} {\bf 76} (1996) pp. 348-351 
%
\bibitem{1goueld}
Eldredge N. and Gould S. J., in {\em Models in Paleobiology}, ed. by
T.J.M. Schopf T. J. M., (Freeman, San Francisco, 1972)
%
\bibitem{1ele}
Elena S.E., Cooper V.S. and Lenski R.E., Punctuated evolution 
caused by selection of rare beneficial mutations, {\em Science} {\bf 272}
(1996) pp. 1802-1804 and p. 1741
%
\bibitem{2fg}
Fussy S. and Gr\"ossing G., Fractal
evolution of normalized feedback systems on a lattice,
{\em Phys. Lett.} {\bf A186} (1994) pp. 145-151, nlin.AO/0204047 
%
\bibitem{2fgs1}
Fussy S., Gr\"ossing G. and Schwabl H., 
Fractal evolution in deterministic and random models,
{\em Int. J. Bifurcation and Chaos} {\bf 6}(11) (1996) pp. 1977-1995
%
\bibitem{2fgs2}
Fussy S., Gr\"ossing G. and Schwabl H., 
Fractal evolution in discretized systems. In 
{\em Self-Organization of Complex Structures: From Individual to Collective
Dynamics}, ed. by Schweitzer F. (Gordon and Breach, London, 1996) in press 
%
\bibitem{2fgs3}
Fussy S., Gr\"ossing G. and Schwabl H., 
Hierarchically emergent fractal evolution.
In {\em Cybernetics \& Systems '96 Vol. I},
ed. by Trappl R. (Vienna, 1996) pp.189-194
%
\bibitem{2goueld}
Gould S. J. and Eldredge N., Punctuated equilibrium comes of age, {\em Nature}
(1993) {\bf 366} pp. 223-227 
%
\bibitem{1kauff}
Kauffman S. A., {\em The origins of Order}
(Oxford Univ. Press, New York Oxford, 1993)
%
\bibitem{lens}
Lenski R. E. and Traviso M., Dynamics of adaptation and diversification: A
10,000-generation experiment with bacterial populations,
{\em Proc. Natl. Acad. Sci.} {\bf 91} (1994) pp. 6808-6814 
%
\bibitem{mandel}
Mandelbrot B. B., {\em The fractal geometry of nature}
(W. H. Freeman, San Francisco, 1982) pp.286-287
%
\bibitem{schust}
Schuster P., Molekulare Evolution an der Schwelle zwischen Chemie und
Biologie. In {\em Die Evolution der Evolutionstheorie}, ed. by Wieser W.
(Spektrum Akademie Verlag, Heidelberg-Berlin-Oxford, 1994) pp.49-76
%
\bibitem{snepp}
Sneppen K., Bak P., Flyvbjerg H. and Jensen M.H., Evolution as a
self-organized critical phenomenon, 
{\em Proc. Natl. Acad. Sci.} {\bf 92} (1995) pp. 5209-5213
%
\bibitem{sole1}
Sol\'e R. V. and Manrubia S. C., Extinction and self-organized criticality
in a model of large-scale evolution, {\em Phys. Rev.} {\bf E54} (1996) pp. 
R42-R45
%
\bibitem{vand}
Vandewalle N. and Ausloos M., Self-organized criticality in 
phylogenetic-like tree growths, {\em J. Phys. I France} {\bf 5} (1995)
pp. 1011-1025
%
\end{thebibliography}
\end{document}